\documentclass[preprint,showpacs,preprintnumbers,amsmath,amssymb]{revtex4}

\usepackage{graphicx}% Include figure files
\usepackage{dcolumn}% Align table columns on decimal point
\usepackage{bm}% bold math

\begin{document}
\preprint{APS/123-QED}

\title{Stochastic Particle Acceleration in Multiple Magnetic Islands during Reconnection}% Force line breaks with \\

\author{Masahiro Hoshino}
 \email{hoshino@eps.s.u-tokyo.ac.jp}
 \affiliation{Department of Earth and Planetary Science, University of Tokyo, Tokyo 113-0033, Japan}

\date{\today}% It is always \today, today,
% but any date may be explicitly specified

\begin{abstract}
A nonthermal particle acceleration mechanism involving the interaction of a charged particle with multiple magnetic islands is proposed.  The original Fermi acceleration model, which assumes randomly distributed magnetic clouds moving at random velocity $V_c$ in the interstellar medium, is known to be of second-order acceleration of $O(V_c/c)^2$ owing to the combination of head-on and head--tail collisions.  In this letter, we reconsider the original Fermi model by introducing multiple magnetic islands during reconnection instead of magnetic clouds.  We discuss that the energetic particles have a tendency to be distributed outside the magnetic islands, and they mainly interact with reconnection outflow jets.  As a result, the acceleration efficiency becomes first-order of $O(V_A/c)$, where $V_A$ and $c$ are the Alfv\'en velocity and the speed of light, respectively.  

\end{abstract}

%PACS numbers may be entered using the \verb+\pacs{#1}+ command.
\maketitle

%%% introduction %%%
Particle acceleration during magnetic reconnection is a fundamental problem in collisionless plasma, and it is now widely recognized that magnetic reconnection is one of the important nonthermal particle acceleration processes in our universe.  The nonthermal particles observed in solar flares and substorms in the Earth's magnetotail are known to be driven by reconnection \cite{lin03,oieroset02}, and magnetic reconnection now receives a great deal of interest in many astrophysical settings such as pulsars, magnetars, galaxy clusters, and active galactic nucleus jets \cite{kirk04,zenitani07}.

There has been much acclaimed research into particle acceleration through reconnection.  The most elementary process is meandering/Speiser acceleration in the diffusion region with a weak magnetic field \cite{speiser65,pritchett05}.  The charged particles inside the diffusion region can be directly accelerated almost along the reconnection electric field until they escape the diffusion region.  However, the diffusion region is very small compared with the total system of reconnection, and the amount of energetic particle flux produced by this process is thus not necessarily enough to explain the observation of nonthermal energy density, which seems to account for at least several tens of percent of the total energy density \cite{lin03}.

Another mechanism is betatron-type acceleration around the pileup magnetic field region where Alfv\'enic outflow jets 
in the reconnecting magnetic field 
collide with preexisting plasma at rest \cite{hoshino05,imada07}.  In the pileup region, the plasma can gain energy proportional to the magnitude of the compressed magnetic field.  In addition to the betatron-type acceleration, if the gyroradius of a charged particle is of the order of the magnetic field curvature radius, the chaotic particle behavior can enhance the particle energization \cite{buchner89}.  Additionally, the electromagnetic waves generated by reconnection can be an agent of the nonadiabatic process \cite{hoshino05}.

Another important acceleration process can occur for reconnection with multiple magnetic islands.  For a long current sheet, many elongated magnetic islands can form from tearing or plasmoid instability \cite{biskamp86}.  An elongated magnetic island then starts to shrink, and the particles trapped inside the island can be accelerated through a Fermi-type process \cite{drake06b}.  If the gyroradius of the accelerated particles exceeds the size of the island, the particles escape the island and their energization ceases.

Turbulent magnetic reconnection has been also discussed as a possible mechanism of particle acceleration \cite{ambrosiano88}.  The magnetic tension force of a tangled magnetic field line in turbulent reconnection is suggested to be as an important agent of particle acceleration \cite{lazarian09}.  However, the acceleration efficiency of this process remains an open question.

In this letter, we investigate particle energization in the interaction of a charge particle with many magnetic islands, and propose a new acceleration process extended from the original Fermi acceleration mechanism \cite{fermi49}.  In the original Fermi acceleration mechanism, particles gain energy stochastically during head-on and head--tail collisions of particles with magnetic clouds or mirrors as the scattering objects.  During the stochastic acceleration, the increase in particle energy is known to be second order of $V_c/c$ ($V_c$ and $c$ are the velocity of the random motion of the magnetic cloud and the speed of light, respectively), and is apparently a slow process.  
Here we study the dynamic evolution of many magnetic reconnection sites instead of magnetic clouds, and we show that the energetic particles have a tendency to interact with the reconnection outflow jets. The acceleration efficiency is thus strongly enhanced relative to that of the original Fermi acceleration, and the energy increase becomes first order of $V_A/c$, where $V_A$ is the Alfv\'en velocity.

%%%%% figure 1
%----------------------------------------------------------------------------
\begin{figure*}
\includegraphics[width=16.4cm]{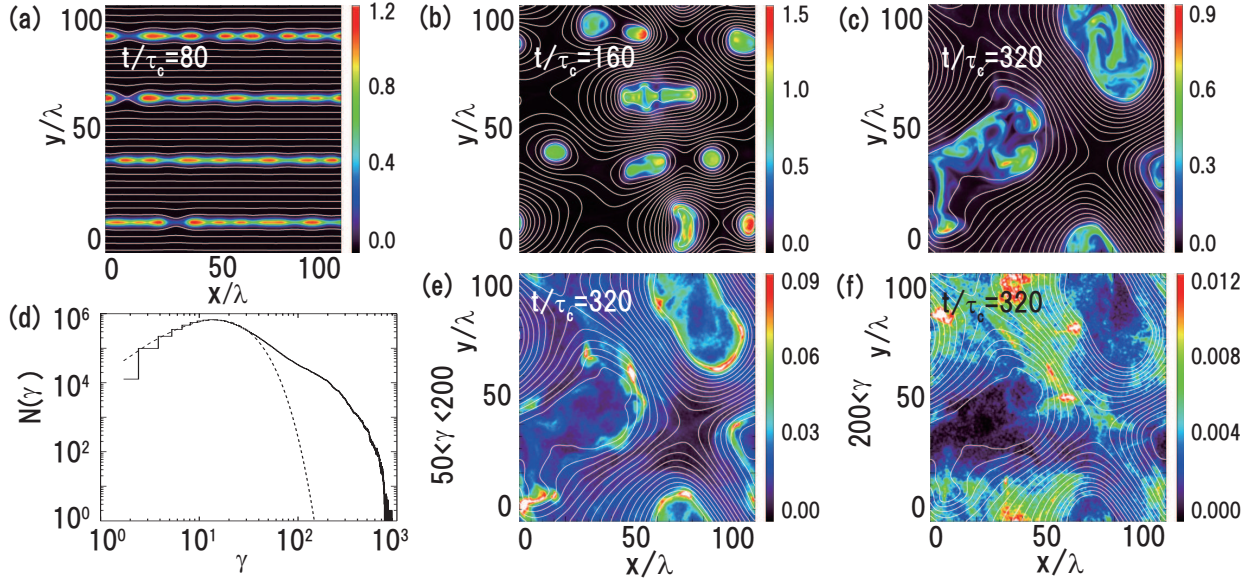}
\caption{Time evolution of multiple magnetic islands. Panels (a--c) show the plasma density (color contour) and magnetic field lines (white), and panel d shows the energy spectrum at $t/\tau_c=320$. The dashed line is the Maxwellian fit.  
Panels (e and f) show the densities for the mid-energy and high-energy particles, respectively.  }
\end{figure*}
%----------------------------------------------------------------------------

%%% simulation %%%
To study the stochastic acceleration during the interaction with magnetic islands, simulations are carried out with a relativistic particle-in-cell (PIC) code.  Shown in Figure 1 is the time evolution of the magnetic islands and the energy spectrum.
As the initial condition, four Harris current sheets of the relativistic pair plasmas with peak density $n_0$ superimposed on uniform background density $n_b=0.05n_0$ are set up in two-dimensional space of $x \times y=512 \times 512$ girds.  
The initial plasma temperature is assumed to be $T_0/mc^2=5$ for both electron and positron, and the drift velocities of electrons and positrons are $v_d/c=\pm0.5$, respectively.  
The plasma skin depth $c/\Omega_p=1.12 \Delta$, the gyroradius $c/\Omega_c=0.25 \Delta$, and the magnetization parameter $(\Omega_c/\Omega_p)^2=20$, where the grid size $\Delta$.
The total number of particles is $512 \times 512 \times 200$.  The double-periodic boundary condition is assumed for both $x$ and $y$ directions.  A set of Maxwell's equations are solved in the Fourier space so that the numerical dispersion error can be suppressed.  
The simulation results are presented in normalized units: the magnetic field is normalized to the asymptotic value of the antiparallel component of the Harris model $B_0$, the density to $n_0$, length to the initial plasma sheet thickness $\lambda = 5 \Delta$, and time to the light transit time $\tau_c=\lambda/c$.  A uniform guide magnetic field $B_z=B_0$ is added.

Shown in Figure 1a--c is the time evolution of the plasma density (color contour) and the magnetic field lines (white).  In the early stage of the linear tearing mode instability ($t/\tau_c=80$ in panel a), approximately eight magnetic islands form in each current sheet, the time evolution of which is consistent with our previous simulation \cite{zenitani07}.  As time goes on (b and c), the magnetic islands grow by coalescing and merging, and the islands from adjacent current layers overlap.  At $t/\tau_c=320$ (panel c), two large magnetic islands containing most of the plasma density are seen.

During the above evolution, many nonthermal particles are generated.  Figure 1d shows the energy spectrum, where the horizontal and vertical axes are the Lorentz factor $\gamma$ of the particle energy and the number density $N(\gamma)$.  The solid line is the energy spectrum at $t/\tau_c=320$, and the dashed line is a Maxwellian spectrum fitted by $T/mc^2 =7.3$.  
During the evolution of reconnection, the initial thermal plasmas are heated and accelerated.  
There is clear formation of nonthermal particles above $\gamma > 40$, and the maximum energy can reach $\gamma \sim 10^3$, for which the gyroradius is almost half the size of the simulation system.  The nonthermal particle occupies about 66 \% of the total particle energy.

How and where energetic particles are accelerated is now investigated.  We plot the densities for the energetic particles in panels e and f of Figure 1.  We calculate the densities by integrating the velocity distribution function with the particle momentum according to $\int_{p_{min}}^{p_{max}} f(p) dp$.  Panels e and f show the densities of the energetic populations with $(p_{min},p_{max})=(50,200)$ and $(200,\infty)$, respectively.  ($(p_{min},p_{max})= (0,\infty)$ for panels a--c. )  The gyroradius for the energetic particle with $p/mc=200$ is about $10 \lambda$, which roughly corresponds to the size of the magnetic island during the island merging stage.  In contrast to the total plasma density, we see that (1) the middle energy range (panel e) is localized around the magnetic island, and (2) the high-energy plasma (panel f) forms a void structure inside the islands.  We find that the thermal plasma is mainly confined in the islands (i.e., the weak magnetic field region), while the energetic particles are preferentially distributed outside the islands (i.e., the stronger magnetic field region).  

%%%% figure 2
%----------------------------------------------------------------------------
\begin{figure}
\includegraphics[width=8.0cm]{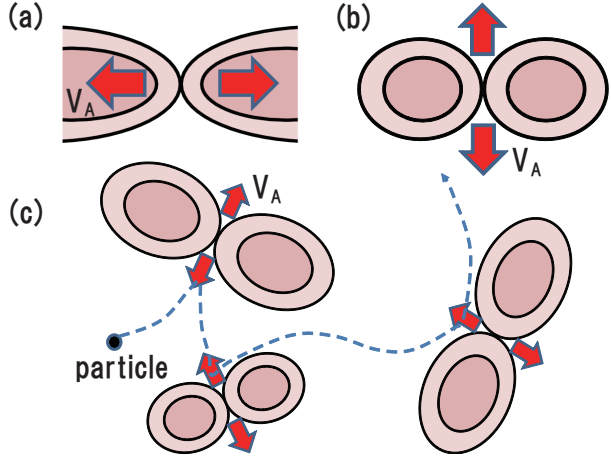}
\caption{Illustration of (a) the plasma sheet reconnection in the early stage, (b) the reconnection during magnetic island merging/coalescence in the late stage, and (c) the interaction of an energetic particle with magnetic islands.  Black lines show magnetic field lines, red arrows show the reconnection outflows with Alfv\'en speed $V_A$, and the blue dashed line is the trajectory of the energetic particle.}
\end{figure}
%----------------------------------------------------------------------------

The reason why energetic particles have the tendency to be localized in the strong magnetic field is simply illustrated in Figure 2. Reddish regions surrounded by black lines are the magnetic islands and red arrows are the Alfv\'enic reconnection jets.  In the early evolution of magnetic reconnection (panel a), the energetic particles are known to be generated in and around the X-type reconnection region, and they are ejected into the plasma sheet in association with the reconnection outflow jet \cite{speiser65}.  On the other hand, in the later phase of magnetic coalescence/island merging (panel b), the reconnection outflows are always directly towards the strong magnetic field region \cite{oka10}.  That is, the thermal plasma is supplied by the two merging magnetic islands, and energetic particles are generated in the current sheet sandwiched by two magnetic islands and are ejected into the strong magnetic field region.  The nature of particle energization during the merging of the islands explains the localization of energetic parties in the strong magnetic field region seen in Figure 1f. 

%%%% figure 3
%----------------------------------------------------------------------------
%\begin{turnpage}
%\begin{figure}
%\includegraphics[width=23.0cm]{Figure3_MH.eps}
\begin{figure*}
\includegraphics[width=16.4cm]{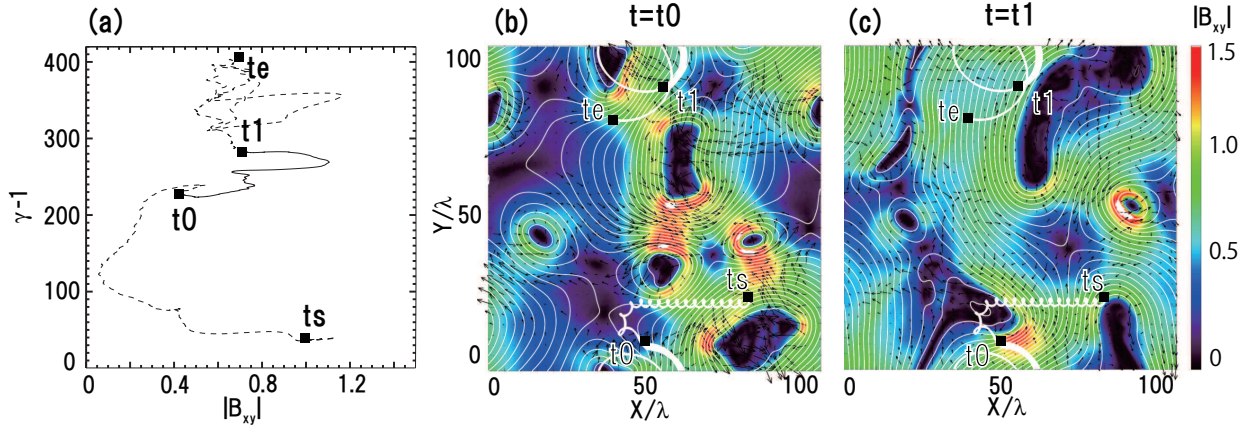}
\caption{A typical particle trajectory during multiple magnetic reconnections.  The particle energization history (a), and the particle trajectories superposed on the magnetic field structures at $t/\tau_c=180$ (b) and at $t/\tau_c=200$ (c).  Color contours show the magnitude of the magnetic field $(B_x^2+B_y^2)^{1/2}$, and black arrows and thin white lines are the plasma flow vector and the magnetic field lines in the reconnection plane.}
\end{figure*}
%\end{figure}
%\end{turnpage}
%----------------------------------------------------------------------------

To study the mechanism of particle acceleration in the strong magnetic field region in detail, we plot a typical particle trajectory and its energy history as a function of the magnetic intensity in Figure 3.  Figure 3a shows the energy history from the initial state at $t_s=0$ to the end of the simulation run at $t_e=320 \tau_c$.  The horizontal and vertical axes are the magnitude of the magnetic field ($B_{xy}= (B_x^2+B_y^2)^{1/2})$ and the particle energy $(\gamma-1)$ normalized by the rest-mass energy, respectively.  At the initial time $t=0$, the particle is located around ($B_{xy}, \varepsilon) = (1,30)$, which is denoted $t=t_s$.  As time passes, the particle gains energy by moving around the reconnection region.  
Note that, roughly speaking, $B_{xy} \le 0.4$ is inside the magnetic island, while $B_{xy} \ge 0.6$ is outside the magnetic island and within the strong magnetic field.  
In the early evolution, the particle is accelerated in the relatively weak magnetic field regions with $B_{xy} \le 0.3$, and we checked that the acceleration occurred in and around the magnetic diffusion region.  This is meandering/Speiser acceleration \cite{speiser65}.  After the initial acceleration in and around the diffusion region, the accelerated particle is ejected from the plasma sheet into the strong magnetic field region during the island merging process, and it moves around the strong magnetic field.  After a while, an accelerated particle encounters another magnetic reconnection region in the island merging stage.  

Figures 3b and 3c show the interaction of a charged energetic particle with the reconnection jet of randomly moving magnetic islands.  The center panel is the particle trajectory superposed on the magnetic field line at $t_0=180 \tau_c$, and the right-hand panel is that at $t_1=200 \tau_c$.  We see that the charged particle can gain energy during the interaction of the charged particle and the reconnection jet with a relatively strong magnetic field of $B_{xy} \sim 0.8$.  After $t>t_1$ the particle can further gain energy in the strong magnetic field region.  The trajectory shows clearly that particle energization occurs during the interaction of the particle and the magnetized plasma flow of the reconnection region.

%%%% mechanism
Based on the above simulation result, we propose a new particle acceleration mechanism as shown in Figure 2c.  
In contrast with the original Fermi acceleration mechanism as a stochastic means by which charged particles collide with magnetic clouds or mirrors \cite{fermi49}, we introduce magnetic islands instead of magnetic clouds or mirrors.  During the merging of magnetic islands, energetic particles have the tendency to be distributed outside the magnetic islands, and the particles collide stochastically with magnetic islands from the outside.  Those particles interacting with the reconnection outflow jets with Alfv\'en velocity $V_A$ will gain energy, and those particles involved in head--tail collisions with the magnetic islands may lose energy.  However, the random velocity $V_c$ may be lower than the thermal velocity $V_s$.  If $V_c > V_s$, the plasma will be heated by a shock wave until $V_c < V_s$ is satisfied.  Therefore, for an active reconnection region where the magnetic field energy dominates the system (i.e., the plasma $\beta < 1$), we may assume $V_c < V_s < V_A$, and the energy loss due to the head-tail collision may be negligible.  The detailed physics of the random velocity remains an open issue.

In our simple particle acceleration model, we did not specify the scattering process in detail, because if we could find a Lorentz transformation into the de Hoffman--Taylor frame during the scattering of a particle, the particle energy gain does not depend on the local scattering process.  If the particles can be reflected back in Alfv\'enic jets, the particle energy gain ($\delta \varepsilon$) is proportional to the Alfv\'en speed; i.e., $\delta \varepsilon/\varepsilon \propto V_A/c$.

In this letter, we presented only the result for the relativistic plasma case, but we confirmed that nonrelativistic reconnection has also the same behavior.  The reconnection outflow velocity does not depend on the relativistic or nonrelativistic plasmas, and the outflow always has the Alfv\'en speed $V_A$.  Once energetic particles are ejected from magnetic islands, our stochastic reconnection process performs well.  The efficiency of energetic particle ejection from the islands (i.e., the so-called injection problem) is an important consideration of this process, and it seems to depend on the plasma dynamics inside the merging current sheet.

The spatial size of the simulations presented here is small compared with any natural system, but we also confirmed in another simulation run that maximum attainable energy was about four times larger for a four times larger simulation box. The maximum attainable energy $\varepsilon_{max}$ can be simply expressed by 
%\begin{equation}
$\gamma_{max}=\varepsilon_{max}/mc^2=eBL/2mc^2 \sim 10^3 (L/100 \lambda)$, 
%\end{equation}
where $L$ is the simulation box size.  This estimation agrees well with Figure 1f.  
This mechanism naturally predicts a power law energy spectrum by taking account of particle loss from the system, but the power law index may depend on the dynamics of the multiple reconnection and the process of particle loss from the active region of island merging.  
To save the computational time, pair plasma has been assumed, but it would be interesting to compare the particle acceleration efficiency for the proton and electron in future study.

Particle acceleration in multiple magnetic islands is important in many astrophysical settings.  
For example, the pulsar wind is known to form a striped wind structure of the toroidal magnetic field in the equatorial plane, and it has been proposed that magnetic reconnection is an important acceleration process that explains synchrotron radiation from the Crab nebula \cite{kirk04}.  At the boundary between the pulsar wind and the synchrotron nebula, a relativistic perpendicular shock is believed to form \cite{kennel84}.  In addition to the shock acceleration, the generation of multiple magnetic islands may be initiated by plasma compression at the shock front.  Some synchrotron particles may be generated in our multiple magnetic reconnection process.  Additionally, a phenomenon similar to striped wind may occur in the termination shock in our heliosphere \cite{lazarian09,drake10}.  

This work was supported in part by a JSPS Research Grant and by the JAXA Supercomputing Center.  M.H. thanks J. F. Drake, L. Vlahos, A. Lazarian, H. Ji, and M. L. Goldstein for discussions.

%\newpage %Just because of unusual number of tables stacked at end
%\bibliography{mrxf}% Produces the bibliography via BibTeX.

\end{document}